\documentclass[aps,prb,onecolumn,floats,superscriptaddress]{revtex4}
\usepackage[pdftex]{graphicx} 
\usepackage{epstopdf}
\usepackage{verbatim}
\usepackage{color}
\usepackage{subfigure}
\usepackage{tabularx}
\usepackage{amsfonts}
\usepackage{wasysym}
\usepackage{bm}
\usepackage{txfonts}
\usepackage{amssymb}
\usepackage{graphicx}
\usepackage{epstopdf}
\usepackage{multirow}
\usepackage{float}
\usepackage[linkcolor=black,anchorcolor=black,citecolor=black,hypertexnames=false]{hyperref}
\usepackage{multibib}
\usepackage{ulem}

\renewcommand\figurename{\textbf{Figure}}

\makeatletter
\def\bib@device#1#2{}
\makeatother

\newcommand{\bra}[1]{\ensuremath{\langle#1|}}
\newcommand{\ket}[1]{\ensuremath{|#1\rangle}}

\begin{document}

\title{Fractionalized excitations in the partially magnetized spin liquid candidate YbMgGaO$_4$}

\author{Yao Shen}
\affiliation{State Key Laboratory of Surface Physics and Department of Physics, Fudan University, Shanghai 200433, China}
\author{Yao-Dong Li}
\affiliation{State Key Laboratory of Surface Physics and Department of Physics, Fudan University, Shanghai 200433, China}
\affiliation{Center for Field Theory and Particle Physics, Fudan University, Shanghai, 200433, China}

\author{H. C. Walker}
\affiliation{ISIS Facility, Rutherford Appleton Laboratory, STFC, Chilton, Didcot, Oxon OX11 0QX, United Kingdom}
\author{P. Steffens}
\affiliation{Institut Laue-Langevin, 71 Avenue des Martyrs, 38042 Grenoble Cedex 9, France}
\author{M. Boehm}
\affiliation{Institut Laue-Langevin, 71 Avenue des Martyrs, 38042 Grenoble Cedex 9, France}

\author{Xiaowen Zhang}
\affiliation{State Key Laboratory of Surface Physics and Department of Physics, Fudan University, Shanghai 200433, China}
\author{Shoudong Shen}
\affiliation{State Key Laboratory of Surface Physics and Department of Physics, Fudan University, Shanghai 200433, China}
\author{Hongliang Wo}
\affiliation{State Key Laboratory of Surface Physics and Department of Physics, Fudan University, Shanghai 200433, China}

\author{Gang Chen$^\ast$}
\affiliation{State Key Laboratory of Surface Physics and Department of Physics, Fudan University, Shanghai 200433, China}
\affiliation{Center for Field Theory and Particle Physics, Fudan University, Shanghai, 200433, China}
\affiliation{Collaborative Innovation Center of Advanced Microstructures, Nanjing, 210093, China}

\author{Jun Zhao$^\ast$}
\affiliation{State Key Laboratory of Surface Physics and Department of Physics, Fudan University, Shanghai 200433, China}
\affiliation{Collaborative Innovation Center of Advanced Microstructures, Nanjing, 210093, China}

\begin{abstract}
\end{abstract}

\maketitle
\textbf{Quantum spin liquids (QSLs) are exotic states of matter characterized by emergent gauge structures and fractionalized elementary excitations. The recently discovered triangular lattice antiferromagnet YbMgGaO$_4$ is a promising QSL candidate, and the nature of its ground state is still under debate. Here, we use neutron scattering to study the spin excitations in YbMgGaO$_4$ under various magnetic fields. Our data reveal a dispersive spin excitation continuum with clear upper and lower excitation edges under a weak magnetic field ($H=2.5$ T). Moreover, a spectral crossing emerges at the $\Gamma$ point at the Zeeman-split energy. The corresponding redistribution of the spectral weight and its field-dependent evolution are consistent with the theoretical prediction based on the inter-band and intra-band spinon particle-hole excitations associated with the Zeeman-split spinon bands, implying the presence of fractionalized excitations and spinon Fermi surfaces in the partially magnetized YbMgGaO$_4$.}

In magnetically ordered Mott insulators, the elementary excitations are the spin-wave-like magnon modes and carry integer spin quantum numbers. In quantum spin liquids (QSLs) that was first proposed by P.W. Anderson as a disordered spin state, however, the situation is drastically different \cite{Anderson1,Anderson2}. A QSL does not have any long-range magnetic order and is an exotic quantum state of matter with long-range quantum entanglement \cite{Balents1,Balents2,PALee1,Norman,ZhouYi}. The description of the QSLs goes beyond the traditional Landau's paradigm that defines phases from their symmetry breaking patterns. For example, the well-known ferromagnets differ from the paramagnets by breaking the time reversal and spin rotational symmetry. In contrast, QSLs are often characterized by the emergent gauge structure with deconfined spinon excitations that carry fractionalized spin quantum numbers \cite{XGWen}. Depending on the type of QSL ground states, the statistics of the spinon excitations can vary from boson to fermion,
or even anyon \cite{Read,Affleck,Balents2,XGWen}. Therefore, convincingly revealing the spinon excitations in a QSL candidate not only confirms the spin quantum number fractionalization, but also provides an important clue about the type of a QSL \cite{Balents1,XGWen}.

\begin{figure}[ht]
\includegraphics[width=1\textwidth]{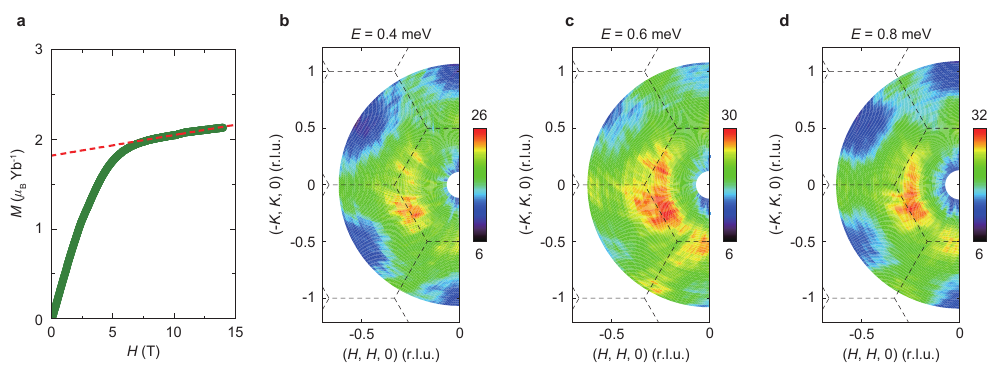}
\caption{ \textbf{Magnetization and spin excitations of YbMgGaO$_4$ under magnetic fields.} \textbf{a,} Magnetic field dependence of the magnetization at \textit{T}=2 K (ref. \onlinecite{Yao}). The dashed line denotes the linear fit of the magnetization above $\sim 7$ T. \textbf{b-d,} Constant-energy images at the indicated energies under a magnetic field of 2.5 T at 70 mK. The colour bar indicates scattering intensity in arbitrary unit in linear scale. The dashed lines represent the Brillouin zone boundaries. The data were collected using the Flatcone detector on ThALES and corrected for neutron-beam self attenuation using a similar method described in ref. \onlinecite{Yao}.
}
\end{figure}

In most cases, the spin quantum number fractionalization and the spinon excitations can be tested by a combination of experimental tools that include thermodynamic, thermal transport, and spectroscopic measurements \cite{Yamashita1,Yamashita2,Yamashita3}. The current experimental study of this question is sometimes constrained by various practical issues, and the experimental confirmation of QSLs remains to be controversial. Indeed, the spinon-like continuum has been observed in some of the QSL candidates \cite{YoungLee,Balz,Yao,Martin2}, and more schemes are needed to provide robust evidence for the very existence of spin quantum number fractionalization \cite{Balents1}. The recent discovery of the triangular-lattice single-crystalline QSL candidate material YbMgGaO$_4$ provides a new testing ground for the QSL research \cite{Yuesheng1}. No ordering or symmetry breaking is observed down to about 30 mK in a variety of measurements \cite{Yao,Martin2,Yuesheng1,Yuesheng3,Yuesheng4,XuYang}. It is argued that the spin-orbital entanglement induced anisotropic interactions may trigger strong quantum fluctuations and help stabilize the QSL state consequently \cite{Yuesheng2,Yaodong1,polarized}.
More substantially, recent inelastic neutron scattering measurements have discovered a broad spin excitation continuum covering a large portion of the Brillouin zone \cite{Yao,Martin2,Yuesheng5}. A variety of theoretical proposals have been made including the QSL state with a spinon Fermi surface and nearest-neighbour resonating valence bonds state \cite{Yao,Yuesheng5,Yaodong2,Yaodong3}. Meanwhile, the scenario of disorder and spin-glass state have also been recently suggested \cite{Sasha, Wen}. To further confirm the fermionic spinon excitation and the spin quantum number fractionalization in YbMgGaO$_4$, the field dependent evolution of the spin excitations needs to be tested \cite{Yaodong4,oleg}. In fact, the fermionic spinon excitation in one spatial dimension has been examined in this manner. For the spin-1/2 Heisenberg chain, that is essentially a one dimensional Luttinger liquid, and is not a true QSL in the modern sense, but shows spinon excitations in the form of domain walls, the external magnetic field could lead to the splitting of the spinon band, resulting in a modulation and redistribution of the spinon continuum \cite{Dender,Halg}. These results provided a firm confirmation of the fractionalized spinon excitation in one spatial dimension.

In this paper, we use inelastic neutron scattering technique to study the spin excitations of YbMgGaO$_4$ single crystals under various external magnetic fields. In a weak magnetic field of $H=2.5$ T applied along the $c$ axis, a dispersive spin excitation continuum is revealed with clear upper and lower excitations edges, leading to a spectral crossing at the $\Gamma$ point at the Zeeman-split energy. The corresponding redistribution of the spectral weight and its field-dependent evolution are inconsistent with the conventional magnon behavior, but instead are unique features for fractionalized spinon excitations. To be more specific, we show that the measured dynamic spin structure is consistent with the theoretical prediction based on the inter-band and intra-band spinon particle-hole excitations associated with the Zeeman-split spinon bands. Our results provide an important piece of evidence for fractionalized spinon excitations and spinon Fermi surfaces in YbMgGaO$_4$ under magnetic fields.

\begin{figure}[ht]
\includegraphics[width=1\textwidth]{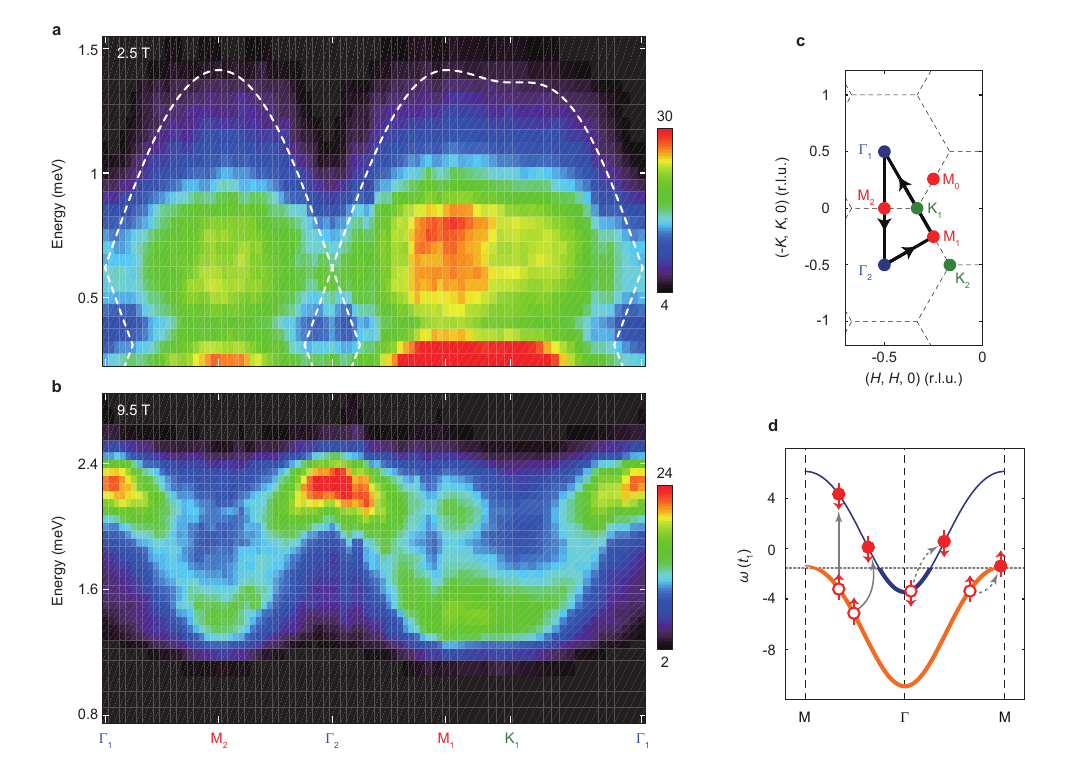}
\caption{ \textbf{Intensity of the spin excitation spectrum along the high symmetry directions at 2.5 T and 9.5 T at 70 mK.} \textbf{a,} Contour plot of the energy dependent intensity at 2.5 T along the high symmetry directions illustrated by the black lines in \textbf{c}. The white dashed lines indicate the calculated boundaries of the continuum based on the spinon Fermi surface model described in Methods. \textbf{b,} Contour plot of the energy dependent intensity in the nearly polarized state at 9.5 T. The colour bar indicates scattering intensity in arbitrary unit in linear scale. \textbf{c,} Sketch of reciprocal space. The dashed lines indicate the Brillouin zone boundaries. \textbf{d,} The split spinon band structure along the high-symmetry points (vertical dashed lines) in momentum space. $t_1$ is the nearest-neighbour spinon hopping. The blue and orange bands are of spin-down and spin-up spinons, respectively. The horizontal dotted line indicates the Fermi level. The solid arrows indicate the spin-flipped inter-band particle-hole excitations while the dotted arrows indicate spin-unflipped intra-band particle-hole excitations.
}
\end{figure}

\textbf{Results}

\textbf{Continuous excitations under weak external fields.} We start by examining the magnetization of a YbMgGaO$_4$ single crystal under magnetic fields along the \textit{c} axis (Fig. 1a). The magnetization increases progressively with field below $\sim 4$ T, followed by a smooth transition into the high-field regime above $\sim 7$ T, where the magnetization nearly saturates. This is consistent with previous measurements \cite{Yuesheng2}. To study the effect of magnetic field on the continuous spin excitations in YbMgGaO$_4$, we first use inelastic neutron scattering to measure the spin excitations under $H = 2.5$ T in the low field regime.  As can be seen in Fig. 1b-d, the constant-energy images show ring-like shaped diffusive magnetic excitations covering a wide region of the Brillouin zone. Such momentum structure of spin excitations resembles the continuous excitations observed at zero field, which was interpreted as the evidence for spinon excitations \cite{Yao,Martin2}. This implies that the spinon excitations persist under a weak magnetic field, as we will discuss subsequently.

To determine the dispersion of the spin excitation continuum in the weak field regime, we measured the energy dependence of the spectral intensity along the high symmetry directions (Fig. 2a). Unlike the zero-field data in which the spectral weight near the zone center ($\Gamma$ point) is strongly suppressed, a prominent enhancement can be seen at $\sim 0.6$ meV, leading to a spectral crossing near $\Gamma$. Moreover, the continuum shows clear lower and upper excitation edges near $\sim 0.6$ meV (marked by the X-shaped cross in Fig. 2a) around $\Gamma$, which is distinct from the V-shaped upper excitation edge at zero field \cite{Yao}. As the energy is further lowered, there is another upper
excitation edge below $\sim$ 0.3-0.4 meV (marked by the V-shaped edge in Fig. 2a). The dispersion of the broad continuum is further confirmed by the constant energy cuts along the high-symmetry directions in Fig. 3.

\begin{figure}[ht]
\includegraphics[width=0.8\textwidth]{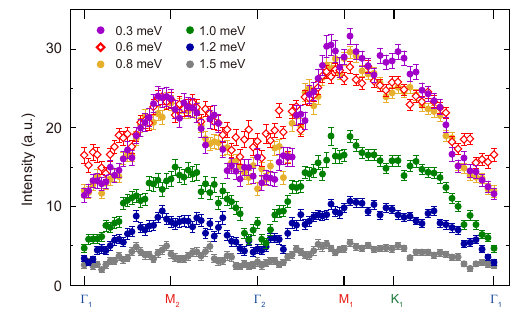}
\caption{ \textbf{Constant energy scans at 2.5 T and 70 mK.} The cuts are made along the high symmetry directions, $\Gamma$-M-$\Gamma$-M-K-$\Gamma$, at the indicated energies. Error bars, 1 s.d.; a.u., arbitrary unit.
}
\end{figure}

\textbf{Field dependence of the spin excitations.}To gain further insights into the origin of the continuum, we present in Fig. 4 the field and energy dependence of the spin excitation at $\Gamma$, M and K points. At the zero field, it is found that the spectral weights mainly spread along the zone boundary, resulting in a suppressed intensity near the zone centers ($\Gamma$ points). At low fields, however, a spectral peak occurs at a finite energy at $\Gamma$ (Fig. 4a), which corresponds to the spectral crossing point in Fig. 2a. As the field is increased, the spectral peak shifted to higher energy in a linear manner as denoted in Fig. 4b. Meanwhile, the broad continuum at M and K persists except that the spectral intensity is gradually suppressed with increasing field. It seems that part of the spectral weight has been transferred from M and K to $\Gamma$ point and no clear shift of the overall continuum is detected. Such behavior differs from what one would expect for the spin-wave excitations in a conventional magnet, where the whole spin wave band should shift to high energy with increasing field, since magnons couple to magnetic field directly \cite{Yaodong1,Martin1}. Indeed, sharp spin-wave excitations with a $\sim 1.2$ meV gap are observed under a high magnetic field of 9.5 T in a nearly polarized state (Fig. 2b). Moreover, the high-field spin-wave spectrum shows a clearly distinct dispersion from that in the low field regime (Fig. 2a). This further indicates that the low-field continuum cannot be magnon excitations.

\textbf{Discussion}

We propose that the modulation of the spectral weights of the continuum in the low field regime is consistent with the previously predicted behavior of the spinon Fermi surface QSL state under magnetic fields \cite{Yaodong4}. In the weak field regime, the proposed zero-field spinon Fermi surface QSL state is expected to persist and the spinon remains to be a valid description of the magnetic excitation \cite{Yaodong4}, which is confirmed by our data that continuum excitations are observed at all energy measured. It was previously shown in ref. \onlinecite{Yaodong4} that, the degenerate spinon bands are split and the splitting is given by the Zeeman energy. The mean-field results for the specific parameter choice of the present experiment are given in details in Methods. In an inelastic neutron scattering measurement, the neutron energy-momentum loss creates the spin excitation that at the mean-field level corresponds to both the inter-band and intra-band particle-hole excitation of the spinons. The particle-hole excitation continuum of the spinons persists into the weak field regime. In particular, for zero momentum transfer of the neutron, the relevant particle-hole excitation would simply be the vertical inter-band excitation between the spin-up and spin-down spinon bands and leads to the spectral peak at the $\Gamma$ point and the Zeeman-split energy (Fig. 2a, 2d and 4a). For momenta away from the $\Gamma$ point, depending on how the momenta are connected to the two split spinon bands, the inter-band particle-hole continuum is bounded by the upper and lower excitation edges that cross each other at the Zeeman-split energy and the $\Gamma$ point (marked by the X-shaped excitation edges in Fig. 2a). As for the intra-band particle-hole excitations below the Zeeman-split energy, a minimal momentum transfer, $p_{min} \approx E/v_{\rm{F}}$, is needed to create the intra-band particle-hole excitation across each spinon Fermi surface for a small and finite neutron energy loss $E$, where $v_{\rm{F}}$ refers to the corresponding Fermi velocity. Thus, the spinon continuum near the $\Gamma$ point is bounded by an upper excitation edge (marked by the V-shaped edge below the X-shaped edge in Fig. 2a).

\begin{figure}[ht]
\includegraphics[width=1\textwidth]{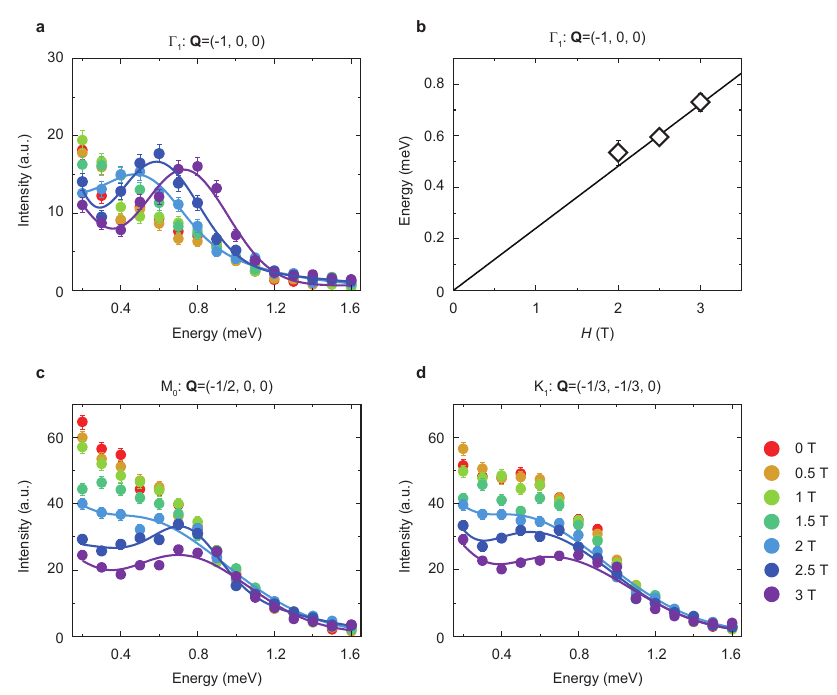}
\caption{ \textbf{Field dependence of constant-Q cuts at high symmetry points at 70 mK.} \textbf{a,} Constant-\textbf{Q} scans at $\Gamma_1$ point under different fields at 70 mK. \textbf{b,} A linear fit to the spectral peak positions presented in \textbf{a} gives the Land\'{e} \textit{g}-factor of 4.16(8). \textbf{c, d,} Constant-\textbf{Q} scans at M$_0$ and K$_1$ points. The solid lines in \textbf{a}, \textbf{c} and \textbf{d} in are guides to the eye.  The high intensity at $\sim$0.2 meV is due to background contamination from incoherent elastic scattering at \textit{E} = 0 meV. The data are collected in the single-detector mode.
}
\end{figure}

It is the spinon Fermi surface and the fractionalized nature of the spinons that give the excitation continuum in the zero field \cite{Yao} and the particular spectral structure of the continuum in the weak field regime. The magnetic field shifts the spin-up and spin-down spinon bands in an opposite fashion. Being fractionalized spin excitations,
the spin-up and spin-down spinons should combine together and contribute to the magnetic excitations measured by the inelastic neutron scattering. This opposite behavior of the two spinon bands in magnetic field is manifested by the spectral peak and the spectral crossing at the $\Gamma$ point as well as the lower and upper excitation edges near the $\Gamma$ point. Indeed, the calculated spinon excitation spectrum associated with the split spinon bands (Supplementary Fig. 2c) captures such features of the measured excitation continuum (Fig. 2a).

Despite a reasonable agreement between the experimental data and the theoretical prediction, we would also point out the discrepancy for a future improvement on the theory side. The spectral peak at the $\Gamma$ point in Fig. 4a shows a wider broadening and a reduced intensity compared to the theoretical expectation \cite{Yaodong4}. This spectral broadening and reduced intensity could arise from two sources. One is the intrinsic broadening due to the gauge fluctuation. The spinon-gauge coupling not only blurs the quasiparticle nature of the spinon but also gives extra scattering processes with gauge photons involved in the inelastic neutron scattering. Both effects are beyond the mean-field analysis for non-interacting spinons and will be discussed in future work. The other is the extrinsic broadening due to the $g$-factor randomness. It was argued that Mg/Ga site disorder may create crystal electric field variations and thus induce $g$-factor randomness \cite{Yuesheng4}. The $g$-factor randomness will give the Zeeman energy disorder and is thus responsible for the broadening of the spectral peak. Finally we emphasize that, the spin quantum number fractionalization with spinon excitations is one of the key properties of QSL and could survive even with weak local perturbations such as weak disorder.

We would also like to discuss about other scenarios. Recently ref. \onlinecite{Yuesheng5} suggested the nearest-neighbor resonating valence bond (RVB) scenario and claimed that the excitation continuum in the inelastic neutron scattering measurement bears no obvious relation to spinons. In fact, it is well-known that the nearest-neighbor RVB state on frustrated lattices such as the triangular lattice is a fully gapped $\mathbb{Z}_2$ QSL \cite{Moessner1, Moessner2, Balents3}. As a result, even in the nearest-neighbor resonating valence bond scenario, the excitation continuum should be the spinon continuum. For such a $\mathbb{Z}_2$ QSL, all the excitations, both spinons and visons, are gapped, and the spinon gap would be of the order of the exchange coupling. There does not seem to be any signature of gapped visons and spinons in the heat capacity and the spin susceptibility data. As was explained in our previous work, the usual gapped $\mathbb{Z}_2$ QSL is a bit difficult to reconcile with the dynamic spin structure factor in this system \cite{Yao}.

Apart from the scenario of the nearest-neighbor RVB state, the scenario with disorders was also proposed \cite{Sasha}. Disorders might play some role in the very low energy magnetic properties and the thermal transport \cite{XuYang,PALee3}. The important questions are whether the disorder is the driving force of the possible QSL state in YbMgGaO$_4$, and to what extent the disorder affects the possible QSL state in YbMgGaO$_4$. Spin quantum number fractionalization, that is one of the key properties of QSL, could persist even with weak disorder. In fact, both the excitation continuum in previous experiments and the redistribution of the continuum under a weak field in the present experiment provide strong support for spin quantum number fractionalization in YbMgGaO$_4$ (Refs. \onlinecite{Martin2,Yao}).

In addition, a spin-glass scenario has been proposed in ref.~\onlinecite{Wen}, where a.c. susceptibility shows a broad peak at an extremely low temperature of $\sim$80-100 mK. The authors attribute this behavior to a spin glass transition. However, it is known that a spin liquid can also display a peak structure in a.c. susceptibility because of the presence of slow dynamics \cite{Balz}. In fact, the spin glass scenario is also in conflict with $\mu$SR measurements where no spin freezing was observed down to 48 mK (Ref. \onlinecite{Yuesheng3}). Moreover, this a.c. susceptibility peak appears at such a low temperature where the $R\ln$2 magnetic entropy has been already released by more than $99\%$. Furthermore, the redistribution of the spin excitation continuum under a weak field observed in our current work is difficult to be explained by the magnon behavior of a spin glass, but can be explained straightforwardly by fractionalized spinon excitations. These suggest that the peak structure of a.c. susceptibility could due to slow dynamics which is the result of the gapless spinon excitations in this material.

\textbf{Methods}

\textbf{Inelastic neutron scattering experiments.} Inelastic neutron scattering measurements were performed on the ThALES cold triple-axis spectrometer at the Institut Laue-Langevin, Grenoble, France. High-quality YbMgGaO$_4$ single crystals were synthesized using the optical floating zone technique\cite{Yao}. PG(002) was used as a monochromator. In the single-detector mode measurements (Fig. 4), PG(002) was used as an analyzer and the final neutron energy was fixed at ${E_\textrm{f} = 3.5}$ meV, resulting in an energy resolution of $\sim$ 0.11 meV. For the measurements with the Flatcone detector (Fig. 1-3, Supplementary Fig. 1), Si (111) was used as an analyzer and the final energy was fixed at ${E_\textrm{f} = 4}$ meV, resulting in a energy resolution of $\sim$0.16 meV. A velocity selector was installed in front of the monochromator to remove the contamination from higher-order neutrons. A dilution insert was used to reach temperatures down to $\sim 70$ mK in the vertical magnet.

In order to reduce the influence of the neutron beam self-attenuation (by the sample), same correction method is used as that in ref. \onlinecite{Yao} for data shown in Fig. 1-3. The self-attenuation effect can be presented as anisotropic intensity distribution in the elastic incoherent scattering image measured at 20 K (Supplementary Fig. 1a). Similar anisotropy is also observed in the raw constant-energy images in the inelastic channel (Supplementary Fig. 1b-d). The correction can be done by normalizing the inelastic data with a linear attenuation correction factor converted from the elastic incoherent scattering intensity which is dependent on the sample position ($\omega$) and scattering angle (2$\theta$). The normalized constant-energy images are presented in Fig. 1b-d. All the data in the manuscript are presented without symmetrization/folding.

\textbf{Spinon Fermi surface in a weak magnetic field.} Here we explore the coupling of the candidate spinon Fermi surface state for YbMgGaO$_4$ to the external magnetic field. This spinon Fermi surface QSL state was originally proposed for the triangular lattice organic materials $\kappa$-(ET)$_2$Cu$_2$(CN)$_3$ and EtMe$_3$Sb[Pd(dmit)$_2$]$_2$  (refs \onlinecite{Shimizu,Itou,PALee2,Motrunich1}). For the organics, due the small Mott gap and proximity to the Mott transition, the coupling to the magnetic field may involve a significant Lorentz coupling \cite{Motrunich2}. For YbMgGaO$_4$ that is in the strong Mott regime, however, only Zeeman coupling is necessary \cite{Yaodong4}.

In previous works, spinon mean-field theory and a systematic projective symmetry analysis have suggested a SU(2) rotational invariant spinon mean-field Hamiltonian with short-range spinon hoppings to describe YbMgGaO$_4$ (refs \onlinecite{Yao,Yaodong3}). A more recent theoretical work by two of us has extensively studied the effect of weak magnetic field on the spinon continuum based on mean-field theory \cite{Yaodong4}. Here we adjust the early theoretical formulation into the parameter choice of the current experiment. In the current experiment, the field is applied along the $c$ axis (normal to the triangular plane). From the SU(2) symmetry of the spinon mean-field theory,
the direction of the magnetic field will probably not induce any qualitatively different behavior on the spinon continuum from the $c$-direction magnetic field at the mean-field level.

Here we explain the basic idea and the underlying physics, and also point out the difference from the zero-field results \cite{Yao,Yaodong3}. We introduce the Abrikosov fermion representation for the spin operator such that ${{\boldsymbol S}_i = \sum_{\alpha\beta}f^{\dagger}_{i\alpha} \frac{{\boldsymbol{\sigma}}_{\alpha\beta}}{2} f^{}_{i\beta}}$ with the Hilbert space constraint ${\sum_{\alpha} f^{\dagger}_{i\alpha} f^{}_{i\alpha} = 1}$. We start with the mean-field Hamiltonian for the spinons,
\begin{eqnarray}
{\mathcal H}_{\rm{MFT}} = -t_1 \sum_{\langle ij \rangle,\alpha}
f^{\dagger}_{i\alpha} f^{}_{j\alpha}-t_2 \sum_{\langle\langle ij \rangle\rangle,\alpha}
f^{\dagger}_{i\alpha} f^{}_{j\alpha}
-\mu \sum_i f^{\dagger}_{i\alpha} f^{}_{i\alpha}
- g_z \mu_{\rm{B}} H \sum_i
f^{\dagger}_{i\alpha} \frac{\sigma^z_{\alpha\beta}}{2}  f^{}_{i\beta}
\end{eqnarray}
where $t_1$ and $t_2$ are the nearest and next-nearest neighbour spinon hoppings, respectively. The chemical potential $\mu$ is introduced to impose the Hilbert space constraint, and the last Zeeman terms accounts for effects of the external magnetic field along the $c$ axis. Since the system is in the strong Mott regime, the charge fluctuation is strongly suppressed, the Lorentz coupling due to charge fluctuation in the weak Mott regime does not apply here \cite{Motrunich2}. We only need to consider the Zeeman coupling to the magnetic field \cite{Yaodong4}. We choose the hopping term in ${\mathcal H}_{\rm{MFT}}$ to be spatially uniform, since it was shown to be the only symmetric mean-field state that is compatible with the existing experiments \cite{Yaodong3}. The fractionalized nature of the spin excitations is already captured by this simple spinon mean-field Hamiltonian, and the further neighbor spinon hopping is introduced to improve the comparison with experiments. We remark on the SU(2) spin rotational symmetry of the spinon mean-field Hamiltonian ${\mathcal H}_{\rm{MFT}}$. This SU(2) spin symmetry at the mean-field level is protected by the projective symmetry group \cite{Yaodong3}. This symmetry is clearly absent in the microscopic spin model \cite{Yaodong1}. It is then pointed out \cite{Yaodong3,Yaodong4} that the anisotropic spin interaction enters as SU(2) symmetry breaking interactions between the spinons. A random phase approximation was then introduced to capture the anisotropic interaction and compute the dynamic spin structure factor. It was found that the spectral weight of the spinon continuum is redistributed and the qualitative features of the continuum persist. More detailed mean-field theory and the random phase approximation have been discussed in the previous theoretical works \cite{Yaodong3,Yaodong4}.

Without the magnetic field, the ground state of Eq. 1 is a filled Fermi sea of degenerate spin-up and spin-down spinons with a large Fermi surface. It has already been shown that, the particle-hole continuum of the spinon Fermi surface gives a consistent explanation for the excitation continuum in the inelastic neutron scattering measurement with the zero field \cite{Yao,Yaodong2}. Moreover, due to the spin rotational invariance and the degenerate spin-up and spin-down spinon bands, the spin-flipping process and the spin-preserving process in the neutron scattering, that correspond to the inter-band particle-hole excitation and the intra-band particle-hole excitation respectively, give the same momentum-energy relation for the inelastic neutron scattering spectrum. Therefore, in the previous calculations \cite{Yaodong3,Yao}, considering the inter-band particle-hole excitation is sufficient.

In the presence of a weak magnetic field $H$, such that the Zeeman coupling would only have a perturbative effect on the QSL ground state and the spinon remains to be a valid description of the magnetic excitation, the previously degenerate spin-up and spin-down bands are now split by an energy separation set by the Zeeman energy $\Delta \equiv g_z \mu_{\rm{B}} H$ (ref. \onlinecite{Yaodong4}). The inelastic neutron scattering measures the correlation function of the spin component that is transverse to the momentum transfer. The dynamic spin structure factor, that is detected by the inelastic neutron scattering, is given by
\begin{eqnarray}
\mathcal{S}(\boldsymbol{p}, E)
    = (\delta_{\mu\nu} -\hat{p}_{\mu} \hat{p}_{\nu})
    \frac{1}{N} \sum_{i,j} e^{i \boldsymbol{p} \cdot (\boldsymbol{r}_i - \boldsymbol{r}_j)}
      \int dt\, e^{-iEt} \bra{\Omega}
      S^{\mu}(t, \boldsymbol{r}_i) S^{\nu}(0, \boldsymbol{r}_j) \ket{\Omega}
\end{eqnarray}
where $\ket{\Omega}$ is the filled Fermi sea ground state of the splitted spinon bands, and $\hat{\boldsymbol p}$ is a unit vector that defines the direction of the
momentum ${\boldsymbol p}$. Both the $S^+$-$S^-$ correlation and the $S^z$-$S^z$ correlation are involved in the above equation. The correlation between $S^z$ and $S^+$ or $S^-$ is vanishing at the mean-field level because the spinon mean-field Hamiltonian still preserves the U(1) spin rotational symmetry around the $z$ axis. To understand these two contributions, we explain their spectroscopic signatures in turns in the following discussion.

In free spinon mean-field theory, the $S^+$-$S^-$ correlation detects the inter-band spinon particle-hole excitation. We have $\mathcal{S}_{+-}(\boldsymbol{p}, E)$,
\begin{eqnarray}
\mathcal{S}_{+-}(\boldsymbol{p}, E)
    =\frac{1}{N} \sum_{i,j} e^{i \boldsymbol{p} \cdot (\boldsymbol{r}_i - \boldsymbol{r}_j)}
     \int dt\, e^{-iEt} \bra{\Omega} S^+(t, \boldsymbol{r}_i) S^-(0, \boldsymbol{r}_j) \ket{\Omega}
    = \sum_n  \delta(E_n - E_0 - E) |\bra{n} S^+_{\boldsymbol{p}} \ket{\Omega}|^2
\end{eqnarray}
where $n$ refers to the intermediate particle-hole excited state. Since this is a spin-flipping process, it naturally probes the inter-band spinon particle-hole continuum. The $\Gamma$ point, with zero momentum transfer of the neutron, simply corresponds to the vertical transition between the spinon bands (Supplementary Fig. 2a). At the mean-field level, the spectral intensity would be proportional to the density of states available for this vertical transition. Due to the large density of states for this vertical transition, there would be a spectral peak at the $\Gamma$ point and the Zeeman-split energy $\Delta$. We, however, expect that the interaction between the spinons would suppress the mean-field spectral intensity. The actual spectral peak may not be quite significant. In any case, a spectral peak at the $\Gamma$ point and the Zeeman-split energy $\Delta$ is observed in Fig. 4a.

A finite momentum transfer, ${\boldsymbol p}$, would probe the tilted particle-hole process between the two spinon bands. For a fixed and small momentum near the $\Gamma$ point, there exists a range of energies that connect two bands. This indicates the presence of the lower and upper excitation edges that define the energy range of the continuous excitations near the $\Gamma$ point. Moreover, these two edges cross each other right at the $\Gamma$ point and the Zeeman-split energy. These features are observed in Supplementary Fig. 2a.

The $S^z$-$S^z$ correlation detects the intra-band spinon particle-hole excitation, and we have
\begin{eqnarray}
\mathcal{S}_{zz}(\boldsymbol{p}, E)
    = \frac{1}{N} \sum_{i,j} e^{i \boldsymbol{p} \cdot (\boldsymbol{r}_i - \boldsymbol{r}_j)}
      \int dt\, e^{-iEt}
      \bra{\Omega} S^z(t, \boldsymbol{r}_i) S^z(0, \boldsymbol{r}_j) \ket{\Omega}
    = \sum_n  \delta(E_n - E_0 - E) |\bra{n} S^z_{\boldsymbol{p}} \ket{\Omega}|^2,
\end{eqnarray}
where we exclude the static ${\boldsymbol{p}}=0$ contribution from the finite magnetization along the $z$ direction in the actual calculation. Like the zero-field case, the particle and the hole in the intra-band process can be excited right next to the Fermi surface (Supplementary Fig. 2b), thus the intra-band particle-hole excitation can have an arbitrarily small energy. The low energy particle-hole continuum of the intra-band process, however, is bounded by an upper excitation edge near the $\Gamma$ point. This is actually analogous to the one in the zero field result. For a small and finite energy transfer $E$ of the neutron, a minimal momentum transfer ${p_{\rm{min}} \approx E/v_{\rm{F}}}$ is needed to excite the spinon particle-hole pair, where the Fermi velocity $v_{\rm{F}}$ depends on the momentum direction and the spin flavor of the spinon Fermi surface.

Having explained the physics of the inter-band and intra-band scattering, we here include both the inter-band process and the intra-band process and compute the dynamic spin structure in Eq. 2. The result is shown in Supplementary Fig. 2c and is reasonably consistent with the experimental one in Fig. 2a. Note here the calculated spectra show lower intensity at low energy compared to the experimental results. The observed high intensity at low energies is mainly due to background contamination from the incoherent elastic scattering at $E = 0$ meV, because the contamination is less significant when the energy resolution is improved from 0.16 meV (Fig. 2a) to 0.11 meV (Fig. 4). Another possible cause is owing to the simplicity of the mean-field theory that neglects the U(1) gauge fluctuation. It is well-known that the gauge fluctuation would enhance the low-energy density of spinon excitations and thus increase the spectral weights at low energies.

\textbf{Data Availability Statement}: The data that support the findings of this study are available from the corresponding author upon reasonable request.

\textbf{Acknowledgements}

We thank C. Broholm, F. C. Zhang, M. Mourigal and A. L. Chernyshev for discussions. This work was supported by the Innovation Program of Shanghai Municipal Education Commission (grant number 2017-01-07-00-07-E00018), the Ministry of Science and Technology of China (Program 973: 2015CB921302), the National Key R\&D Program of the MOST of China (grant number 2016YFA0300203), and the National Natural Science Foundation of China (grant number 91421106). Y.-D.L. and G.C. were supported by the Ministry of Science and Technology of
China with the Grant No.2016YFA0301001, the Start-Up Funds and the Program of First-Class Construction of Fudan University, and the Thousand-Youth-Talent Program of China.

Correspondence and requests for materials should be addressed to J.Z. (zhaoj@fudan.edu.cn) or G.C. (gchen\_physics@fudan.edu.cn).

\newpage
\renewcommand\figurename{\textbf{Supplementary Figure}}
\setcounter{figure}{0}
\setcounter{page}{1}

\begin{center}

Supplementary Information for

\textbf{Fractionalized excitations in the partially magnetized spin liquid candidate YbMgGaO$_4$}

Shen \textit{et al.}

\end{center}

\newpage

\begin{figure*}[!h]
\includegraphics[width=1\textwidth]{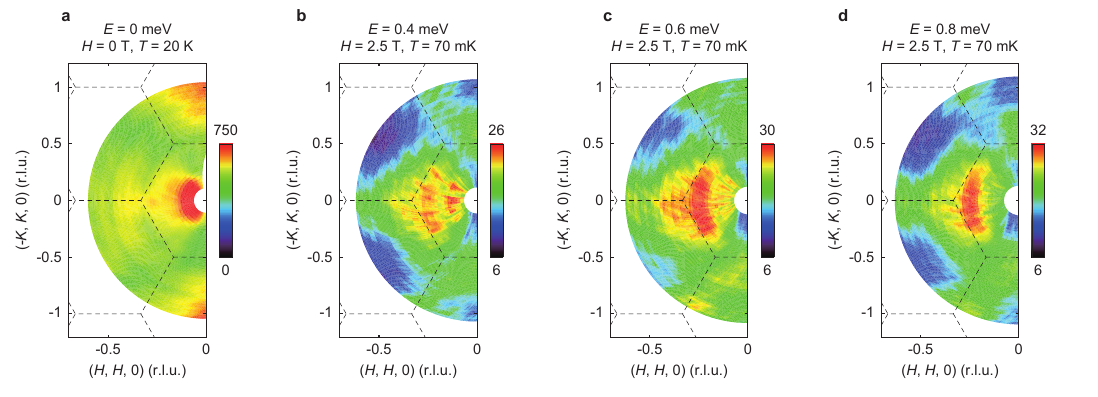}
\caption{ \textbf{Correction of neutron beam self-attenuation.} \textbf{a,} Elastic incoherent scattering image at 20 K and 0 T. \textbf{b-d,} Raw constant-energy images at the indicated energies. Dashed lines indicate the Brillouin zone boundaries. The colour bar indicates scattering intensity in arbitrary unit in linear scale.
}
\end{figure*}

\newpage

\begin{figure*}[!h]
\includegraphics[width=0.8\textwidth]{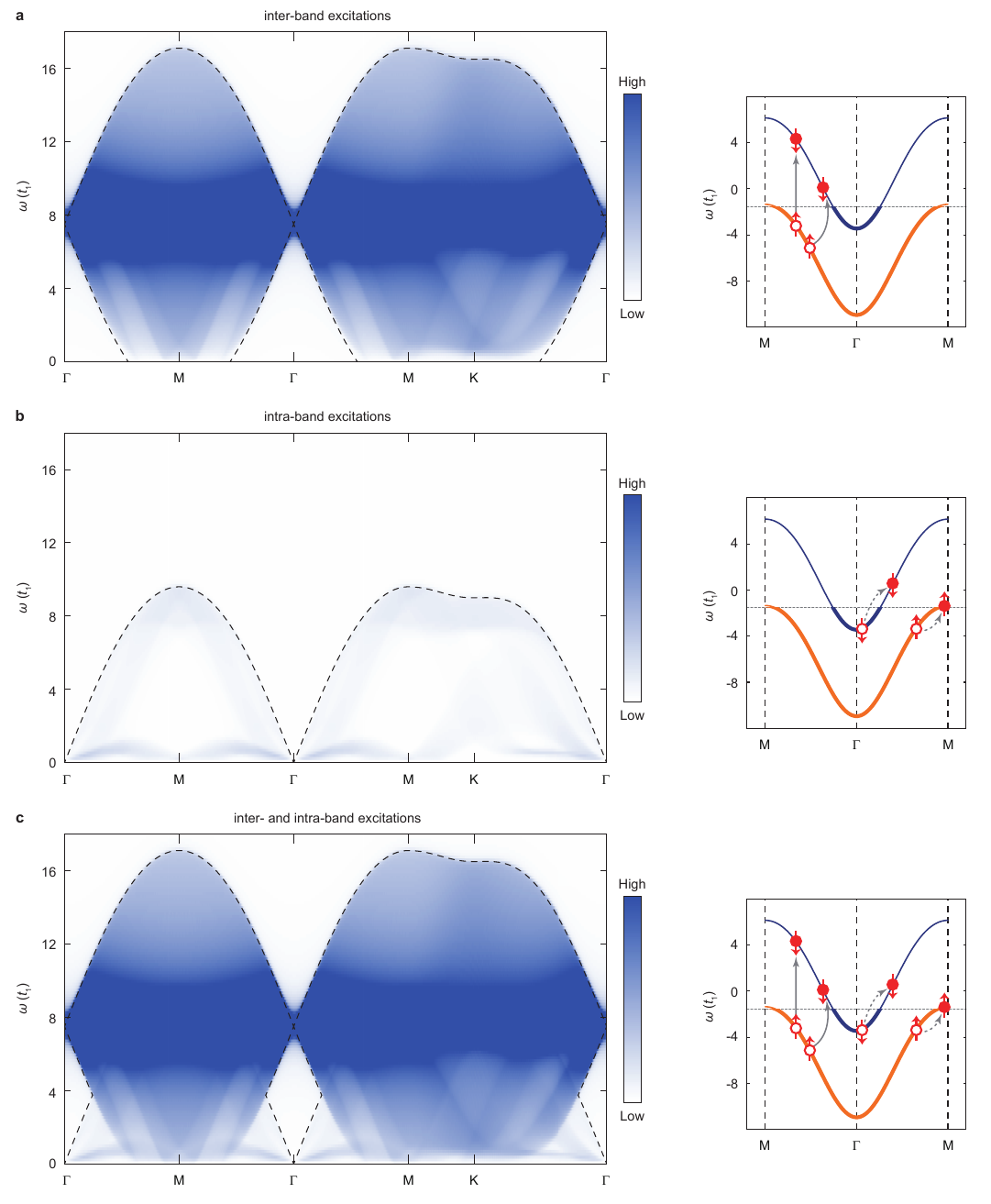}
\caption{ \textbf{Calculation of the spinon band structure and its excitations in a weak magnetic field.} \textbf{a,} The dynamic spin structure factor $\mathcal{S}_{+-}(\boldsymbol{p}, E)$ defined in Eq. 3 (left) in the main text and illustration of the vertical and tilted inter-band particle-hole excitations (right). These are the dominant events that are responsible for the spectral peak at $(\Gamma, \Delta)$ and the upper and lower excitation edges that cross at the peak. \textbf{b,} The dynamic spin structure factor $\mathcal{S}_{zz}(\boldsymbol{p}, E)$ defined in Eq. 4 (left) in the main text and illustration of the intra-band particle-hole excitations (right). These are analogous to the particle-hole excitations in the zero-field case, and also gives rise to an upper excitation edge at $\Gamma$. \textbf{c,} The combined dynamic spin structure factur defined in Eq. 2 in the main that is proportional to the observed neutron scattering density. In all the figures we chose $t_2/t_1$ to be 0.2 and Zeeman splitting gap $\Delta = 7.5t_1$ where $t_1$ and $t_2$ indicate the nearest and next-nearest neighbour spinon hoppings. The colour bar indicates scattering intensity in arbitrary unit in linear scale.
}
\end{figure*}


\textbf{References}
\begin{thebibliography}{}

\bibitem{Anderson1} Anderson, P. W. Resonating valence bonds: a new kind of insulator? \textit{Mater. Res. Bull.} \textbf{8}, 153-160 (1973).

\bibitem{Anderson2} Anderson, P. W. The resonating valence bond state in La$_2$CuO$_4$ and superconductivity. \textit{Science} \textbf{235}, 1196-1198 (1987).

\bibitem{Balents1} Balents, L. Spin liquids in frustrated magnets. \textit{Nature} \textbf{464}, 199-208 (2010).

\bibitem{Balents2} Savary, L. \& Balents, L. Quantum spin liquids: a review. \textit{Rep. Prog. Phys.} \textbf{80}, 016502 (2017).

\bibitem{PALee1} Lee, P. A. An end to the drought of quantum spin liquid. \textit{Science} \textbf{321}, 1306-1307 (2008).

\bibitem{Norman} Norman, M. R. Herbertsmithite and the search for the quantum spin liquid. \textit{Rev. Mod. Phys.} \textbf{88}, 041002 (2016).

\bibitem{ZhouYi} Zhou, Y., Kanoda, K. \& Ng, T.-K. Quantum spin liquid states. \textit{Rev. Mod. Phys.} \textbf{89}, 025003 (2017).

\bibitem{XGWen} Wen, X.-G. Quantum orders and symmetric spin liquids. \textit{Phys. Rev. B} \textbf{65}, 165113 (2002).

\bibitem{Read} Read, N. \& Sachdev, S. Large-\textit{N} expansion for frustrated quantum antiferromagnets. \textit{Phys Rev. Lett.} \textbf{66}, 1773-1776 (1991).

\bibitem{Affleck} Affleck, I. \& Marston, J. B. Large-\textit{n} limit of the Heisenberg-Hubbard model: implications for high-$T_c$ superconductors. \textit{Phys. Rev. B} \textbf{37}, 3774-3777 (1988).

\bibitem{Yamashita1} Yamashita, S. \textit{et al.} Thermodynamic properties of a spin-1/2 spin-liquid state in a $\kappa$-type organic salt. \textit{Nat. Phys.} \textbf{4}, 459-462 (2008).

\bibitem{Yamashita2} Yamashita, M. \textit{et al.} Thermal-transport measurements in a quantum spin-liquid state of the frustrated triangular magnet $\kappa$-(BEDT-TTF)$_2$Cu$_2$(CN)$_3$. \textit{Nat. Phys.} \textbf{5}, 44-47 (2009).

\bibitem{Yamashita3} Yamashita, M. \textit{et al.} Highly mobile gapless excitations in a two-dimensional candidate quantum spin liquid. \textit{Science} \textbf{328}, 1246-1248 (2010).

\bibitem{YoungLee} Han, T.-H. \textit{et al.} Fractionalized excitations in the spin-liquid state of a kagome-lattice antiferromagnet. \textit{Nature} \textbf{492}, 406-410 (2012).

\bibitem{Balz} Balz, C. \textit{et al.} Physical realization of a quantum spin liquid based on a complex frustration mechanism. \textit{Nat. Phys.} \textbf{12}, 942-949 (2016).

\bibitem{Yao} Shen, Y. \textit{et al.} Evidence for a spinon Fermi surface in a triangular-lattice quantum-spin-liquid candidate. \textit{Nature} \textbf{540}, 559-562 (2016).

\bibitem{Martin2} Paddison J. A. M. \textit{et al.} Continuous excitations of the triangular-lattice quantum spin liquid YbMgGaO$_4$. \textit{Nat. Phys.} \textbf{13}, 117-122 (2017).

\bibitem{Yuesheng1} Li, Y. \textit{et al.} Gapless quantum spin liquid ground state in the two-dimensional spin-$1/2$ triangular antiferromagnet YbMgGaO$_4$. \textit{Sci. Rep.} \textbf{5}, 16419 (2015).

\bibitem{Yuesheng3} Li, Y. \textit{et al.} Muon spin relaxation evidence for the U(1) quantum spin-liquid ground state in the triangular antiferromagnet YbMgGaO$_4$. \textit{Phys. Rev. Lett.} \textbf{117}, 097201 (2016).

\bibitem{Yuesheng4} Li, Y. \textit{et al.} Crystalline electric-field randomness in the triangular lattice spin-liquid YbMgGaO$_4$. \textit{Phys. Rev. Lett.} \textbf{118}, 107202 (2017).

\bibitem{XuYang} Xu, Y. \textit{et al.} Absence of magnetic thermal conductivity in the quantum spin-liquid candidate YbMgGaO$_4$. \textit{Phys. Rev. Lett.} \textbf{117}, 267202 (2016).

\bibitem{Yuesheng2} Li, Y. \textit{et al.} Rare-earth triangular lattice spin liquid: a single-crystal study of YbMgGaO$_4$. \textit{ Phys. Rev. Lett.} \textbf{115}, 167203 (2015).

\bibitem{Yaodong1} Li, Y.-D., Wang, X. \& Chen, G. Anisotropic spin model of strong spin-orbit-coupled triangular antiferromagnets. \textit{Phys. Rev. B} \textbf{94}, 035107 (2016).

\bibitem{polarized} T\'{o}th, S., Rolfs, K., Wildes, A. R. \& R\"{u}egg, C. Strong exchange anisotropy in YbMgGaO$_4$ from polarized neutron diffraction. Preprint at http://arxiv.org/abs/1705.05699 (2017).

\bibitem{Yuesheng5} Li, Y. \textit{et al.} Nearest-neighbour resonating valence bonds in YbMgGaO$_4$. \textit{Nat. Commun.} \textbf{8}, 15814 (2017).

\bibitem{Yaodong2} Li, Y.-D., Shen, Y., Zhao, J. \& Chen, G. Effect of spin-orbit coupling on the effective-spin correlation in YbMgGaO$_4$. \textit{Phys. Rev. B} \textbf{97}, 125105 (2018).

\bibitem{Yaodong3} Li, Y.-D., Lu, Y.-M. \& Chen, G. Spinon Fermi surface U(1) spin liquid in a spin-orbit-coupled triangular lattice Mott insulator YbMgGaO$_4$. \textit{Phys. Rev. B} \textbf{96}, 054445 (2017).

\bibitem{Sasha} Zhu, Z., Maksimov, P. A., White, S. R. \& Chernyshev, A. L. Disorder-induced mimicry of a spin liquid in YbMgGaO$_4$. . \textit{Phys Rev. Lett.} \textbf{119}, 157201 (2017).

\bibitem{Wen} Ma, Z. \textit{et al.} Spin-glass ground state in a triangular-lattice compound YbZnGaO$_4$. \textit{Phys. Rev. Lett.} \textbf{120}, 087201 (2018).

\bibitem{Yaodong4} Li, Y.-D. \& Chen, G. Detecting spin fractionalization in a spinon Fermi surface spin liquid. \textit{Phys. Rev. B} \textbf{96}, 075105 (2017).

\bibitem{oleg} Luo, Z.-X., Lake, E., Mei, J.-W. \& Starykh, O. A. Spinon magnetic resonance of quantum spin liquids. \textit{Phys. Rev. Lett.} \textbf{120}, 037204 (2018).

\bibitem{Dender} Dender, D. C., Hammar, P. R., Reich, D. H., Broholm, C. \& Aeppli, G. Direct observation of field-induced incommensurate fluctuations in a one-dimensional \textit{S}=1/2 antiferromagnet. \textit{Phys. Rev. Lett.} \textbf{79}, 1750-1753 (1997).

\bibitem{Halg} H\"{a}lg, M., H\"{u}vonen, D., Butch, N. P., Demmel, F. \& Zheludev, A. Finite-temperature scaling of spin correlations in a partially magnetized Heisenberg \textit{S}=1/2 chain, \textit{Phys. Rev. B} \textbf{92}, 104416 (2015).

\bibitem{Martin1} Mourigal, M. \textit{et al.} Fractional spinon excitations in the quantum Heisenberg antiferromagnetic chain. \textit{Nat. Phys.} \textbf{9}, 435-441 (2013).

\bibitem{Moessner1} Moessner, R. \& Sondhi, S. L. Resonating valence bond phase in the triangular lattice quantum dimer model. \textit{Phys. Rev. Lett.} \textbf{86}, 1881-1884 (2001).

\bibitem{Moessner2} Moessner, R., Sondhi, S. L. \& Fradkin, E. Short-ranged resonating valence bond physics, quantum dimer models, and Ising gauge theories. \textit{Phys. Rev. B} \textbf{65}, 024504 (2001).

\bibitem{Balents3} Balents, L., Fisher, M. P. A. \& Girvin, S. M. Fractionalization in an easy-axis Kagome antiferromagnet. \textit{Phys. Rev. B} \textbf{65}, 224412 (2002).

\bibitem{PALee3} Nave, C. P. \& Lee, P. A. Transport properties of a spinon Fermi surface coupled to a U(1) gauge field. \textit{Phys. Rev. B} \textbf{76}, 235124 (2007).

\bibitem{Shimizu} Shimizu, Y., Miyagawa, K., Kanoda, K., Maesato, M. \& Saito, G. Spin liquid state in an organic Mott insulator with a triangular lattice. \textit{Phys. Rev. Lett.} \textbf{91}, 107001 (2003).

\bibitem{Itou} Itou, T., Oyamada, A., Maegawa, S., Tamura, M. \& Kato, R. Quantum spin liquid in the spin-1/2 triangular antiferromagnet EtMe$_3$Sb[Pd(dmit)$_2$]$_2$. \textit{Phys. Rev. B} \textbf{77}, 104413 (2008).

\bibitem{PALee2} Lee, S.-S. \& Lee, P. A. U(1) gauge theory of the Hubbard model: spin liquid states and possible application to $\kappa$-(BEDT-TTF)$_2$Cu$_2$(CN)$_3$. \textit{Phys. Rev. Lett.} \textbf{95}, 036403 (2005).

\bibitem{Motrunich1} Motrunich, O. I. Variational study of triangular lattice spin-1/2 model with ring exchanges and spin liquid state in $\kappa$-(ET)$_2$Cu$_2$(CN)$_3$. \textit{ Phys. Rev. B} \textbf{72}, 045105 (2005).

\bibitem{Motrunich2} Motrunich, O. I. Orbital magnetic field effects in spin liquid with spinon Fermi sea: possible application to $\kappa$-(ET)$_2$Cu$_2$(CN)$_3$. \textit{Phys. Rev. B} \textbf{73}, 155115 (2006).

\end{thebibliography}
\end{document}